\documentclass[aip,sd,reprint,graphicx]{revtex4-1}

\usepackage{graphicx}
\usepackage{epstopdf}
\usepackage{color}

\begin{document}
			
\title{Superradiant Undulator Radiation for Selective THz Control Experiments at XFELs} 

\author{Takanori Tanikawa}
 \affiliation{European XFEL, Holzkoppel 4, 22869 Schenefeld, Germany}

\author{Suren Karabekyan}
 \affiliation{European XFEL, Holzkoppel 4, 22869 Schenefeld, Germany}

\author{Sergey Kovalev}
 \affiliation{HZDR, Institute for Radiation Physics, Bautzner Landstr. 400, 01328 Dresden}

\author{Sara Casalbuoni}
 \affiliation{Institute for Beam Physics and Technology, Karlsruhe Institute of Technology, P.O. Box 3640, D-76021 Karlsruhe, Germany}

\author{Vivek Asgekar}
 \affiliation{HZDR, Institute for Radiation Physics, Bautzner Landstr. 400, 01328 Dresden}
 \affiliation{Physics Dept., S. P. Pune University, Pune 411 007, India}

\author{Svitozar Serkez}
 \affiliation{European XFEL, Holzkoppel 4, 22869 Schenefeld, Germany}

\author{Michael Gensch}
 \email[]{michael.gensch@tu-berlin.de}
 \affiliation{German Aerospace Center (DLR), Institute of Optical Sensor Systems, Rutherfordstrasse 2, 12489 Berlin, Germany}
 \affiliation{Technische Universit\"{a}t Berlin, Institut f\"{u}r Optik und Atomare Physik, Strasse des 17. Juni, 10623 Berlin, Germany}

\author{Gianluca Geloni}
 \email[]{gianluca.geloni@xfel.eu}
 \affiliation{European XFEL, Holzkoppel 4, 22869 Schenefeld, Germany}

\begin{abstract}
The generation of frequency-tunable, narrow-bandwidth and carrier-envelope-phase stable THz pulses with fields in the MV/cm regime that can be appropriately timed to the femtosecond X-ray pulses from free-electron-lasers is of highest scientific interest. It will enable to follow the electronic and structural dynamics stimulated by (non)linear selective excitations of matter on few femtosecond time and {\AA}ngstrom length scales. In this article, a scheme based on superradiant undulator radiation generated just after the XFEL is proposed. The concept utilizes cutting edge superconducting undulator technology and provides THz pulses in a frequency range between 3 and 30~THz with exceptional THz pulse energies. Relevant aspects for realization and operation are discussed point by point on the example of the European XFEL.
\end{abstract}

\maketitle

\section{\label{sec:uno} Introduction, concept and motivations}
X-ray Free-Electron Lasers (for a recent review on XFELs see e.g.~\cite{XFELS1}) are currently the brightest, tunable sources of short X-ray pulses available for basic scientific research. Among many possible uses, the ultrahigh brightness and the ultrashort duration of these pulses can be exploited in a "pump-probe" configuration, to probe non-equilibrium states of a matter sample that can be excited by previous interaction with terahertz radiation.

The application of high-field THz pulses in pump-probe experiments is manifold and a number of groundbreaking experiments have been performed in recent years~\cite{Kampfrath2013, Nicoletti2016}. One fundamental concept here is the selective excitation of specific low energy degrees of freedom, which is only possible when using appropriately narrow-band, multi-cycle and carrier-envelope phase (CEP) stable THz pulses of sufficient intensity. Successful experimental demonstrations range from THz control of electronic phases in correlated materials~\cite{Hu2014, Mitrano2016}, THz control of magnetic order~\cite{Baierl2016} to the alignment of molecules~\cite{Mohsen2017} and the acceleration of free electrons in a vacuum as well as in materials~\cite{Fruehling2009, Turchinovic2015}. The combination with the ultra-short, intense and widely tunable X-ray pulses from XFELs bears the opportunity to understand and verify important phenomena such as the recently observed evidence for THz induced transient, potentially superconducting phases at critical temperatures above room temperature~\cite{Hu2014}. The required THz pulse energy in such experiments is in the few 10~$\mu$J ranges and the repetition rate of the THz pulses acting as a pump should, in the ideal case, be similar to that of the corresponding probing light pulses. Since the latest generation XFELs under early operation stages or construction, such as the  European XFEL~\cite{Altarelli2011} and the LCLS-II~\cite{Galayda2014, NotaL} are based on superconducting technology, THz pulses are also required with repetition rates between 100~kHz and 4.5~MHz. The generation of high-field THz pulses at such high duty-cycle by means of femtosecond lasers does not allow to achieve the required parameters with respect to pulse energy and narrow-bandwidth~\cite{Green2016}; therefore, superradiant undulator radiation emerges as one promising solution. In this paper, we describe the technological feasibility for implementing a few-period superradiant undulator that fits the particular case of the electron linac of the European XFEL. XFEL-class electron beams are characterized by very high quality in terms of emittance, energy spread, and duration and can, therefore, be advantageously used for the production of superradiant THz radiation with the help of an optimized undulator following the main FEL lines. The concept has been pioneered successfully at the XUV FEL FLASH~\cite{Gensch2008}.

However, the energy of the electron beam at the European XFEL is much higher, and during operations is expected to cover the range between 8.5 and 17.5~GeV. As a result, resonance at low photon energies is only obtained at the cost of an extremely high magnetic field. In this work, we will show that the choice of superconducting technology for the undulator is key at high-electron-energy facilities like the European XFEL. Here we will limit ourselves to consider the case of the highest operation energy foreseen for the European XFEL, 17.5~GeV, and THz pulses starting from a fundamental frequency of 3~THz~\cite{CIT1, CIT2b}. In the following Section~\ref{sec:due}, we will discuss suitable THz undulator parameters and expected performance. We will demonstrate that it is possible to automatically obtain pulses with tens of microjoules energies up to several tens of THz within a 10~$\%$ bandwidth. Since the European XFEL is driven by a high-repetition-rate linac, high-repetition THz pulses with the same pattern as the X-ray pulses, that is 10 trains per second with up to 2700 pulses per train, in about 600~$\mu$s, are obtained. Moreover, given the high quality of the electron beam needed to sustain the FEL instability, we do expect an excellent pulse-to-pulse THz energy stability. The feasibility of the THz superconducting undulator will be discussed in Section~\ref{sect:THzSCU}. There we will show that the peak field needed on-axis for achieving a fundamental frequency of 3~THz at 17.5~GeV electron beam energy is about 7.3~T so that well-proven Nb-Ti technology can be used. In Section~\ref{sec:tre} we will discuss possible positions of our THz source at the European XFEL and the transport of the THz pulses through the optics tunnel, while in Section~\ref{sec:quattro} we will see how synchronization/jitter issues for pump and probe signals can be dealt with.

In short, we propose a robust and cost-effective method for the production of relatively narrow bandwidth, frequency-tunable THz sources for pump-probe (THz and X-ray) experiments at the European XFEL. The scientific community would also welcome the option of controlling the number of THz cycles in the radiation pulse. In order to achieve it, the possibility to power a different number of undulator periods will be studied. Energies of several tens of microjoules starting from a frequency of a few THz up to several tens of THz can be obtained, with excellent pulse-to-pulse energy stability, while issues pertaining transport, synchronization, and jitter are found to be no show-stoppers.

\section{\label{sec:due} THz undulator parameters and expected performance}
The main challenge in the construction of a THz undulator for the production of superradiant THz radiation at the European XFEL is the very high-energy of the electron beam, nominally in the range from 8.5 to 17.5~GeV. In fact, the resonance condition on-axis for the first harmonic reads:

\begin{eqnarray}
\lambda = \lambda_U \frac{1 +K^2/2}{2 \gamma^2}~,
\label{resonance}
\end{eqnarray}

with $\lambda = c/\nu$ the fundamental wavelength, $\lambda_U$ the undulator period, $\gamma$ the relativistic Lorentz factor and~\cite{CGS} $K = e B \lambda_U/(2 \pi m c^2)$ the undulator parameter, where $B$ is the on-axis peak magnetic field, $e$ and $m$ the electron charge and the rest mass, respectively. The undulator parameter can also be written in a form convenient for numerical calculation in SI units as $K= 93.36 B[T] \lambda_U[m]$.  Eq.~(\ref{resonance}) implies larger values of the on-axis peak magnetic field for longer resonance wavelengths. As we will see in Section \ref{sec:tre}, superconducting technology is key to achieve THz pulses with the electron energies typical of the European XFEL.

We will limit our consideration starting from a fundamental frequency of 3~THz, corresponding to about $\lambda = $ 100~$\mu$m, and assuming an electron beam energy fixed throughout the paper, unless otherwise explicitly stated, of 17.5~GeV. If we now fix $\lambda_U = 1$~m in Eq.~(\ref{resonance}), we find that in order to reach a minimal frequency of about 3~THz a value $K = 685$ is needed, corresponding to a maximum field on axis $B \simeq 7.3$~T. As we will see in the next Section, this field allows exploitation of well-proven Nb-Ti technology.

As is well-known \cite{PULSE}, the total pulse energy generated by a single electron bunch $W_{bunch}$ while passing through any kind of radiator depends on the bunch length\cite{NotaT} and follows the relation

\begin{eqnarray}
W_{bunch} = N_e [1 + (N_e-1) |\bar{F}(\nu)|^2] W_0
\label{energy}
\end{eqnarray}

where $N_e$ is the number of electrons in the bunch, $W_0$ the pulse energy emitted by a single electron and $\bar{F}(\nu)$ the Fourier transform of the longitudinal current profile of the bunch, $F(t)$, normalized in such a way that its integral over time is unity:

\begin{eqnarray}
\bar{F}(\nu) = \int_{-\infty}^{\infty} d t ~ F(t) \exp(2 \pi i  \nu t)
\label{Fbunch}
\end{eqnarray}

\begin{figure}
\begin{center}
\includegraphics[trim = 0 0 0 0, clip, width=0.5\textwidth]{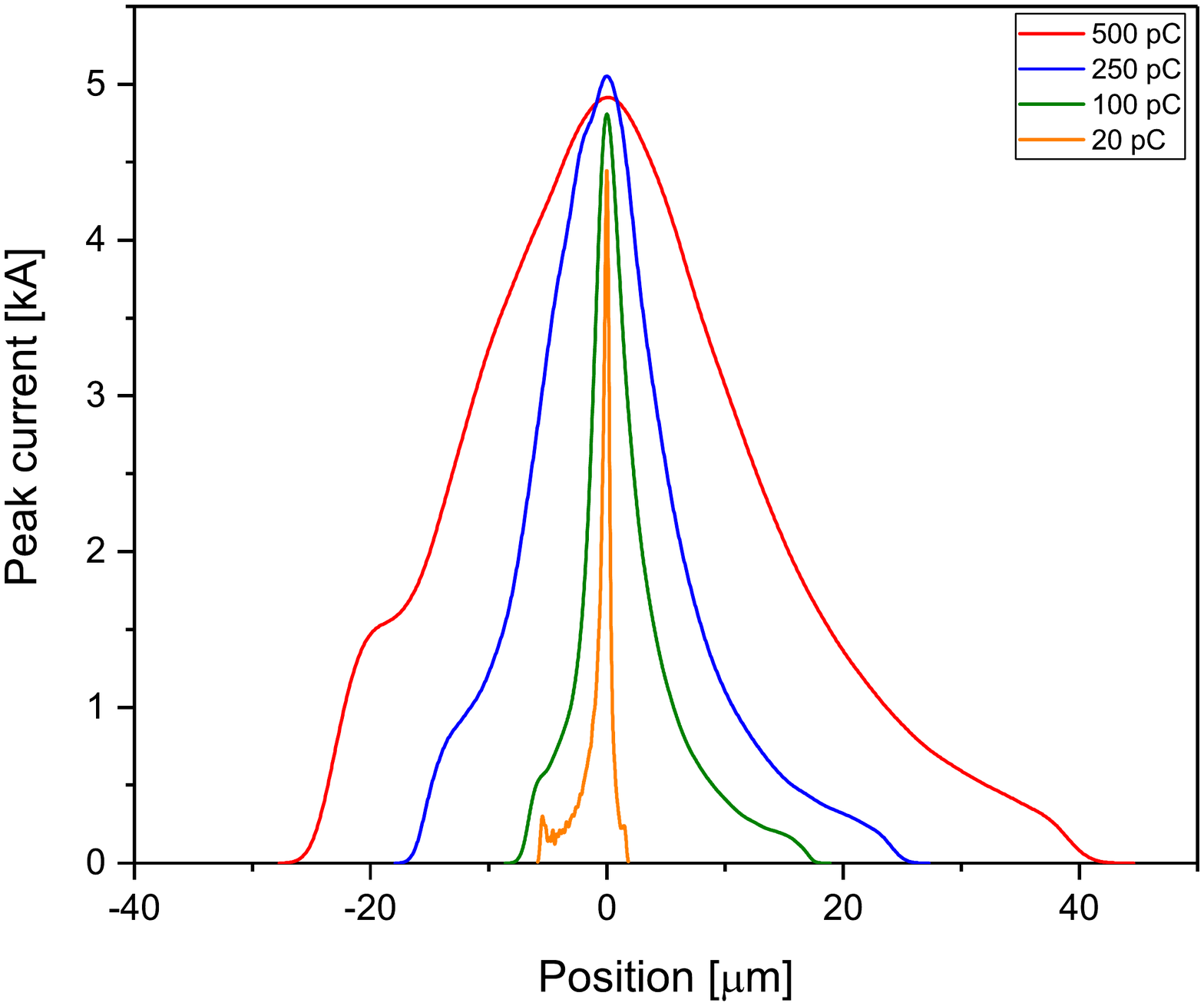}
\end{center}
\caption{Start-to-end simulated electron current profiles for different charges at the European XFEL~\cite{S2E}.}
\label{fig:0}
\end{figure}

The distribution $\bar{F}(\nu)$ is known as a form factor and specifies the degree of superradiance. We see by definition that $\bar{F}(\nu)$ remains roughly near to unity for wavelengths longer than the electron bunch duration, which enables, for those wavelengths, the exploitation of superradiant emission, scaling as $N_e(N_e-1)$. Radiation at shorter wavelengths is, however, strongly suppressed by the form factor. At short wavelengths the usual incoherent emission, scaling as $N_e$, takes place. We conclude that a good electron beam compression is of fundamental importance in achieving superradiant emission at short wavelengths. In order to sustain FEL lasing at X-ray wavelengths, very high peak currents must be achieved (in the order of several kiloamperes for the European XFEL), and hence it is needed to strongly compress the electron bunch in the first place. Therefore, it makes sense to exploit the spent XFEL electron beam for generating THz superradiant pulses, because the form factor is automatically suitable for this goal. The level of compression depends on the charge, with bunches of lower charge being compressed more to obtain the same peak-current level. In Fig.~\ref{fig:0}, the nominal electron current profiles, calculated using start-to-end simulation for the European XFEL with different charges~\cite{S2E}, are reported.

\begin{figure}
\begin{center}
\includegraphics[trim = 0 0 0 0, clip, width=0.5\textwidth]{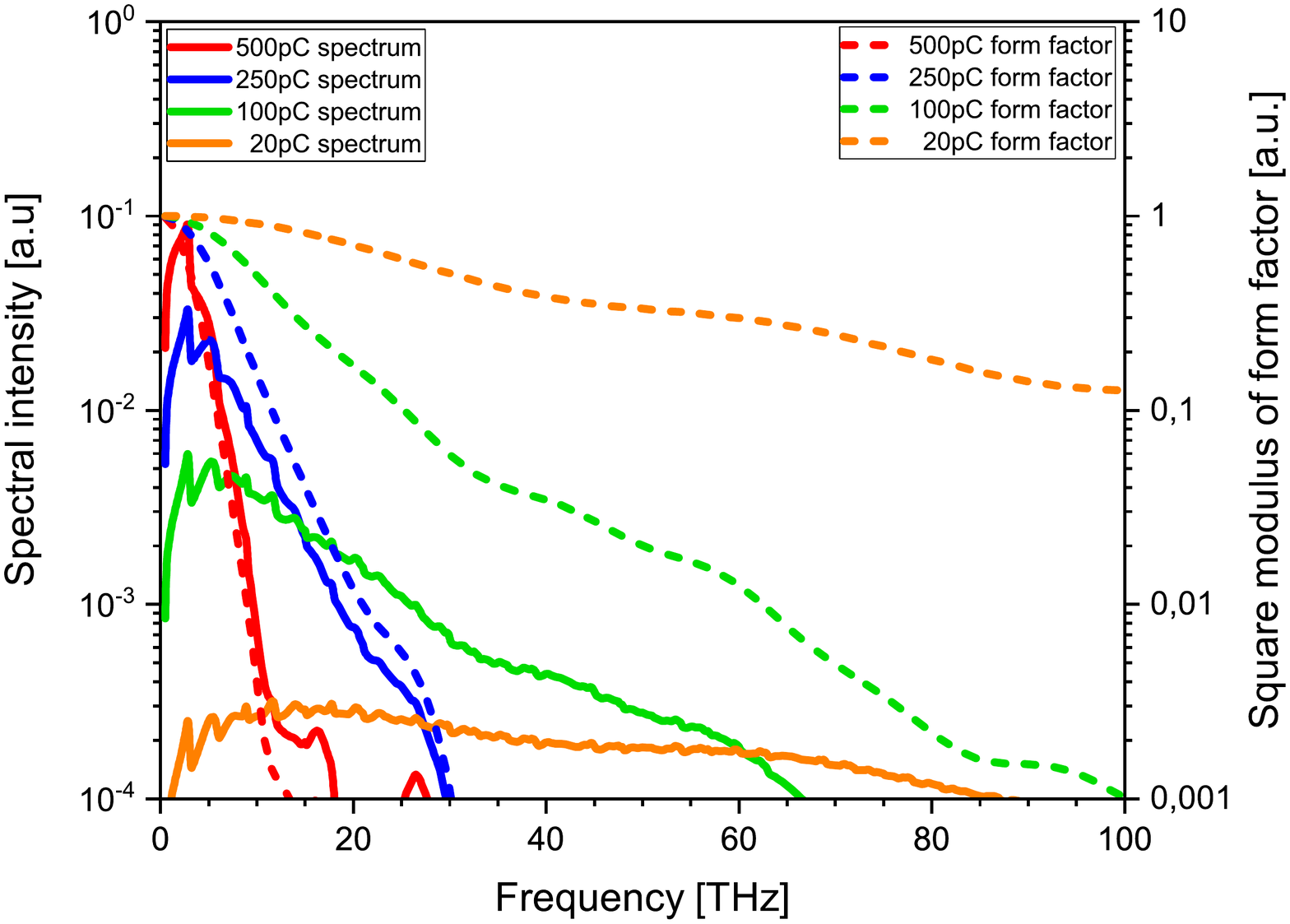}
\end{center}
\caption{Charge dependence of the superradiant THz spectrum at 3~THz fundamental: Square modulus of form factor (dashed lines) and the total fluxes (solid lines).}
\label{fig:2}
\end{figure}

\begin{table}
\caption{Total and fundamental THz pulse energies for different electron charges at the fundamental frequency of 3~THz (using a 100~$\%$ bandpass filter) .}
	\begin{tabular}{c c c}
  \hline
	Electron	&	Total pulse	&	Fundamental	\\
	charge [pC]	&	energy [$\mu$J]	&	pulse energy [$\mu$J]	\\
	\hline
  500	&	3456	&	2058	\\
  250	&	2307	&	750	\\
  100	&	1067	&	136	\\
  20	&	228	&	5.8	\\
	\hline
	\end{tabular}
\label{table:fig2}
\end{table}

In Fig.~\ref{fig:2}, dotted lines, we plot the behavior of the square modulus of the form factor, $|\bar{F}(\nu)|^2$, for different electron bunch nominal charges. The original current profiles were obtained using start-to-end electron beam simulations from the injector of the European XFEL up to the SASE undulators~\cite{S2E}. The FEL process does not change the electron beam current profile on lengths relevant for the production of THz radiation in the undulator so that we can use the same profile at the entrance of the SASE undulators to model the electron beam at the entrance of the THz undulator. The total spectral flux can be obtained by multiplying the square modulus of the form factor by the number of electrons square and by the spectral flux generated by a single electron, which can be easily computed~\cite{NotaS}. Given the very large value of $K$, the angle-integrated spectrum has a very large bandwidth, similar to that of a bending magnet, up to a very high critical frequency that is of no interest here, since the form factor cuts down the superradiant emission much before that point. The result is plotted in Fig.~\ref{fig:2}, solid lines, where we show the total fluxes of the superradiant emission for different charges. The total pulse energies can be calculated by integrating the spectral flux. In our case, they are obtained by integrating up to 100~THz and shown in Table~\ref{table:fig2}.

\begin{figure}
\begin{center}
\includegraphics[trim = 0 0 0 0, clip, width=0.5\textwidth]{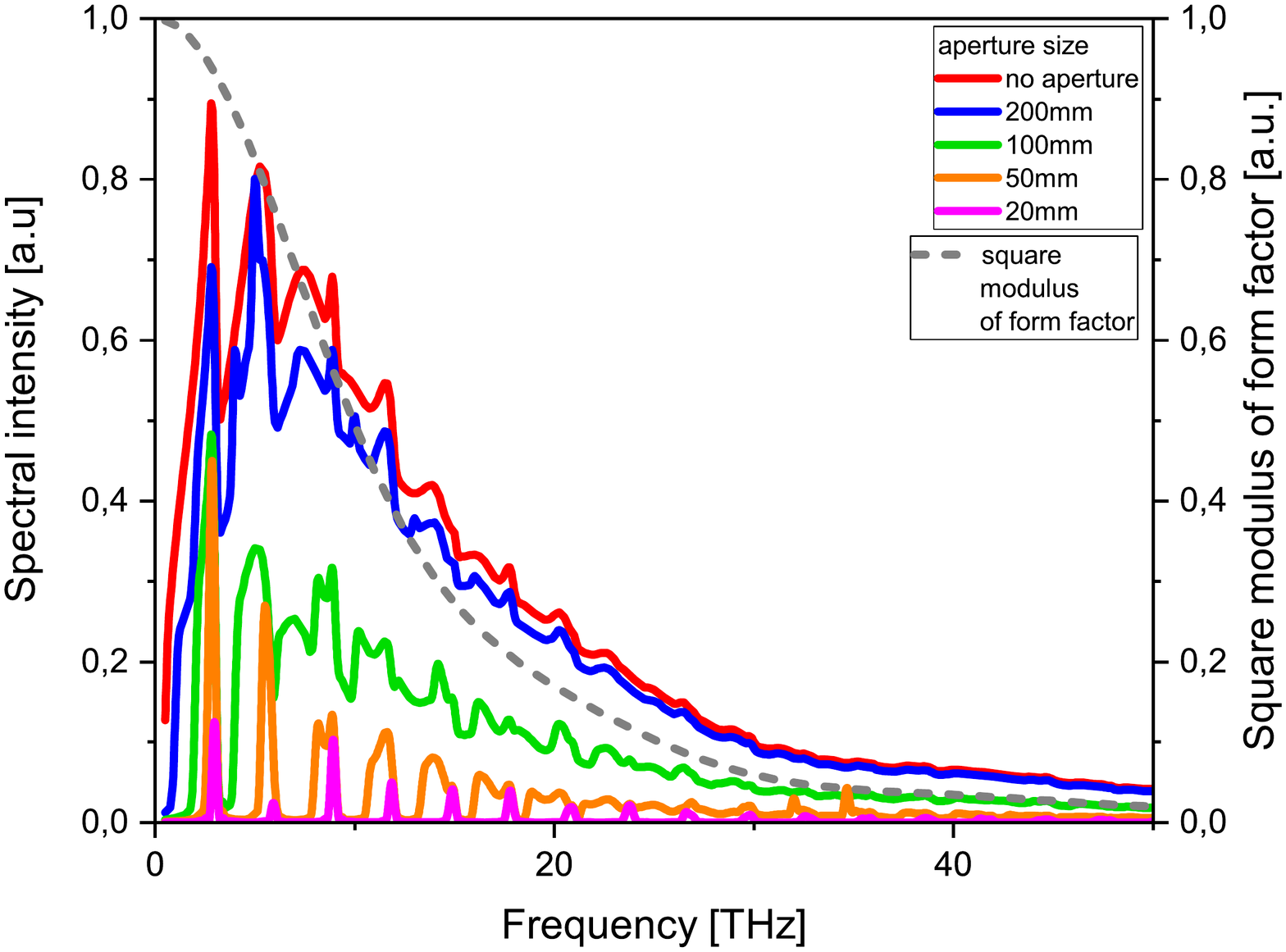}
\end{center}
\caption{Influence of apertures on superradiant THz spectrum at 3~THz fundamental: comparison of THz spectra for the 100~pC case obtained using circular apertures of different diameters around the axis of the system set at 10~m from the center of the THz undulator.}
\label{fig:4}
\end{figure}

\begin{table}
\caption{THz pulse energies and bandwidths upon propagation through different circular apertures at a distance of 10~m from the center of the undulator at the fundamental frequency of 3~THz for the 100~pC case~\cite{noBW}.}
	\begin{tabular}{c c c c}
  \hline
	Aper. size	&	Total pulse	&	Fundamental	&	FWHM Bandwidth of	\\
	{\O} [mm]	&	energy [$\mu$J]	&	pulse energy [$\mu$J]	&	the fundamental [$\%$]	\\
	\hline
  None	&	1067	&	128	&	-	\\
  200	&	914	&	83	&	-	\\
  180	&	851	&	77	&	29	\\
  150	&	722	&	58	&	18	\\
  120	&	541	&	41	&	21	\\
  100	&	403	&	34	&	20	\\
  80	&	272	&	27	&	21	\\
  50	&	107	&	14	&	13	\\
  20	&	16	&	3.0	&	9.1	\\
	\hline
	\end{tabular}
\label{table:fig4}
\end{table}

As shown in Fig.~\ref{fig:4} and in Table~\ref{table:fig4}, by changing the diameter of a circular aperture at a distance of 10~m from the center of the undulator it is possible to select the bandwidth at the different harmonics. By closing the diameter of the circular aperture to 50~mm it is possible to select an FWHM frequency bandwidth of 13~$\%$ of the fundamental while closing it down to 20~mm it is possible to select an FWHM frequency bandwidth of 9.1~$\%$ of the fundamental.

Around the fundamental frequency of 3~THz, we can select a narrow bandwidth using a circular aperture, as can be seen from Fig.~\ref{fig:4}, where we show a comparison of THz spectra for the 100~pC case, obtained using apertures of different diameters set at 10~m from the source, that is in the middle of the undulator. Table~\ref{table:fig4} shows the total pulse energy, the pulse energy under the fundamental and the bandwidth of the fundamental for different aperture sizes.

\begin{figure}
\begin{center}
\includegraphics[trim = 0 0 0 0, clip, width=0.5\textwidth]{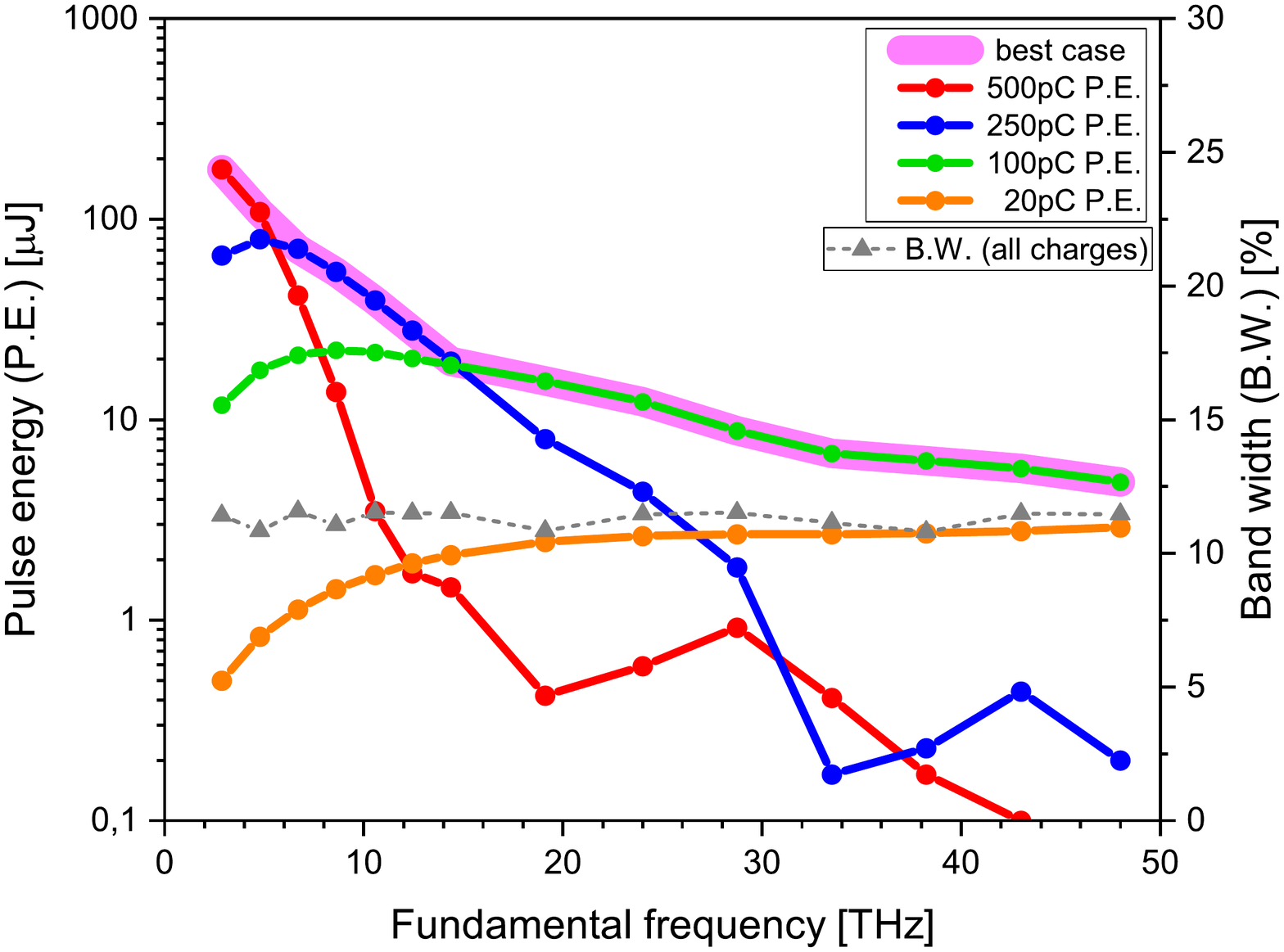}
\end{center}
\caption{Tunability of the THz undulator fundamental: pulse energy at different fundamental frequencies for several electron bunch charges. Here the aperture size is varied so that the bandwidth remains about constant.}
\label{fig:6}
\end{figure}

In Fig.~\ref{fig:6} we show the pulse energy that can be obtained at different fundamental frequencies (obtained by changing the $K$ parameter of the THz undulator) for several nominal electron bunch charges. The aperture size is varied so that the bandwidth remains about constant, around the 10~$\%$ level. From Fig.~\ref{fig:6} we can see a competition between the $N_e^2$ dependence of the pulse energy, which favours high charges, and the form factor near to unity, which favours low charges (at the same peak current, needed to sustain the previous FEL process). As expected, the highest electron charge considered here, 500~pC, yields the best results, in terms of energy per pulse, in the lowest frequency range, between 3 and 6~THz with expected energies above 120~$\mu$J. A charge of 250~pC appears optimal in the range between 6 and 15~THz, with energies above 30~$\mu$J. Finally, the lower charge of 100~pC is suitable for frequencies starting from 15~THz. Our proposed undulator is expected to obtain energies above 10~$\mu$J up to frequencies of about 30~THz. For frequencies higher than that value, the yield drops below the 10$~\mu$J level and remains in the several microjoules level up to frequencies of about 50~THz. Finally, note that the yield of the lowest nominal charge of 20~pC always falls below that of the 100~pC.

\begin{figure}
\begin{center}
\includegraphics[trim = 0 0 0 0, clip, width=0.5\textwidth]{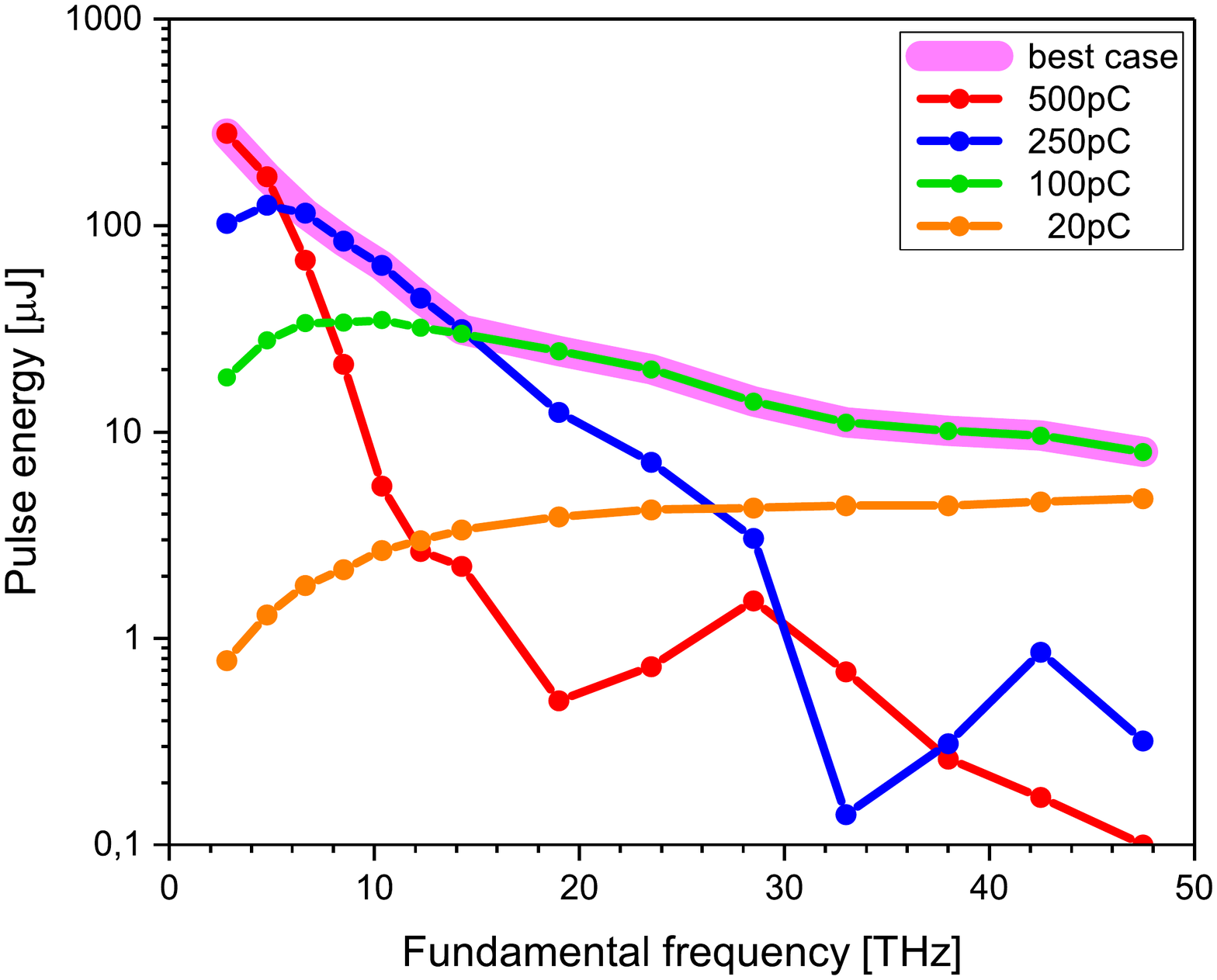}
\end{center}
\caption{Tunability of the THz undulator fundamental: the same as in Fig.~\ref{fig:6} using a 10~$\%$ bandpass filter with boxcar line function instead of apertures.} 
\label{fig:6b}
\end{figure}

We note that the choice of charges is forcefully linked to the operation mode of the XFEL, so that operational restrictions might apply. It should also be remarked that spectral filtering can be used, instead of angular filtering, in order to select a particular spectral range. In Fig.~\ref{fig:6b} we show results of a 10~$\%$ bandpass filtering procedure with a boxcar function. As it can be seen, in this case, larger pulse energies are found.

Photon frequencies lower than 3~THz may, in principle, be reached for lower electron operation energy. For example at 8.5~GeV we could gain about a factor four in the fundamental frequency, pushing the lower limit below 1~THz. Note that while decreasing the electron beam energy only allows us to reach into lower frequencies, leaving the other performance results unvaried. However, assessing the feasibility of this option would require further studies, since a larger undulator gap would be needed to allow for the generation of longer wavelengths without diffraction effects taking place. Moreover, the choice of the electron energy, exactly as the choice of the charge, is obviously linked to the operation mode of the XFEL.

As just mentioned, it is important to look at the expected beam size at the fundamental frequency inside the undulator. This is critical to understand the minimum gap acceptable in the undulator design, under the assumption that we accept the full transverse size at the longest fundamental wavelength. We will consider the case for the fundamental at 3~THz at an electron energy of 17.5~GeV, reminding the reader that a simple scaling for the radiation size $\sigma_{E=E_0 [GeV]}$ applies for a different electron energy $E_0$ expressed in GeV:

\begin{eqnarray}
\sigma_{E=E_0 [GeV]} = \left(\frac{17.5GeV}{E_0 [GeV]}\right) \sigma_{E=17.5 [GeV]}~.
\label{scaling}
\end{eqnarray}

In fact, the square of the electron energy ratio gives the fractional increase in the fundamental frequency that can be reached, while the diffraction size of the source is of order $\sqrt{\lambda L_U/(2 \pi)}$, where $L_U$ is the undulator length.

In order to study the transverse size of the THz source we used two approaches, analytical and numerical. Analytical formulas are available for the field of the radiation source at resonance, in the middle of the undulator, and at any position in free space, after the undulator, even in the near zone~\cite{NEAR}. They read respectively:

\begin{eqnarray}
\widetilde{{E}}(0, r) &=& i \frac{2 \pi K \nu e} {c^2
\gamma} A_{JJ}\left[\pi - 2\mathrm{Si} \left(\frac{2 \pi \nu {r}^2}{L_U c}\right)\right]~, 
\label{undurad5}
\end{eqnarray}

at the source. Here $A_{JJ} = J_0[{K^2}/({4+2K^2})]-J_1[{K^2}/({4+2K^2})]$, $\widetilde{{E}}(0, r)$ is the source field envelope in the frequency domain, $\mathrm{Si}(\cdot)$ is the sin integral function and $r$ is the distance from the $z$ axis on the virtual-source plane (note that $\widetilde{{E}}(0,r)$ is axis-symmetric) and

\begin{eqnarray}
&&\widetilde{E}\left({z},r\right) =\cr &&\frac{2 \pi K \nu e
A_{JJ}}{c^2 \gamma} \left[ \mathrm{Ei} \left(\frac{2 \pi i \nu
r^2}{2{z} c - L_U c}\right)- \mathrm{Ei} \left(\frac{2 \pi i
\nu r^2   }{2{z} c + L_U c}\right) \right]~,  
\label{Esum}
\end{eqnarray}

where $\widetilde{E}\left({z},r\right)$ is the field envelope in the frequency domain at any position $z$ after the undulator, and $\mathrm{Ei}(\cdot)$ is the exponential integral function. Note that Eq.~(\ref{Esum}) is singular at the undulator exit $z = L_U/2$ and for ${r} = 0$. This is related with the use of the resonance approximation.

\begin{figure}
\begin{center}
\includegraphics[trim = 0 0 0 0, clip, width=0.5\textwidth]{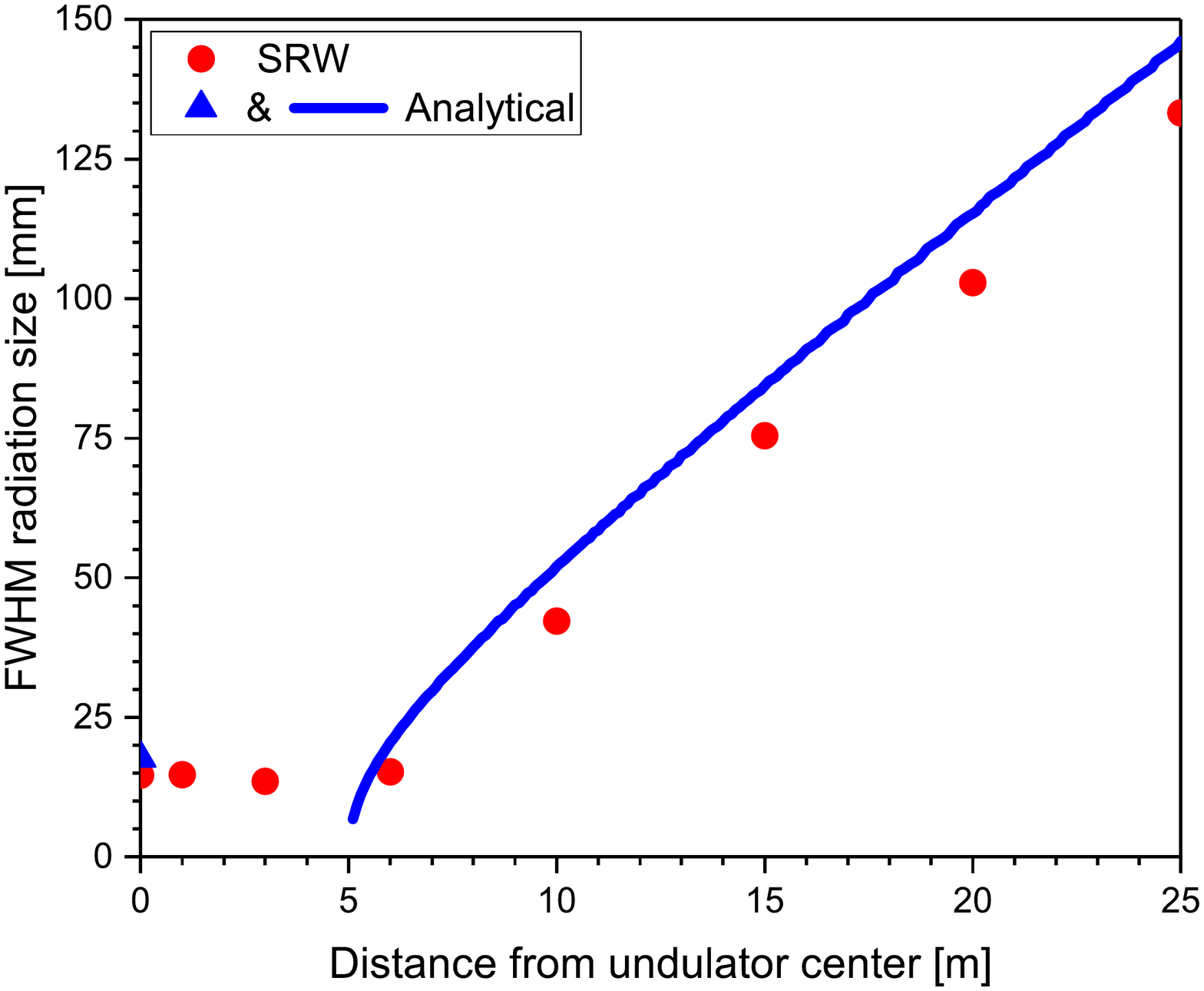}
\end{center}
\caption{Radiation beam size evolution from the undulator center for a fundamental of 3~THz with an electron beam energy of 17.5~GeV. Blue triangles: analytical expectations in the undulator center, at $z=0$ and after the undulator exit according to Eq.~(\ref{undurad5}) and Eq.~(\ref{Esum}) at 3~THz. Red circles: results from wavefront propagation simulations at 3~THz.}
\label{fig:7}
\end{figure}

\begin{figure}
\begin{center}
\includegraphics[trim = 0 0 0 0, clip, width=0.4\textwidth]{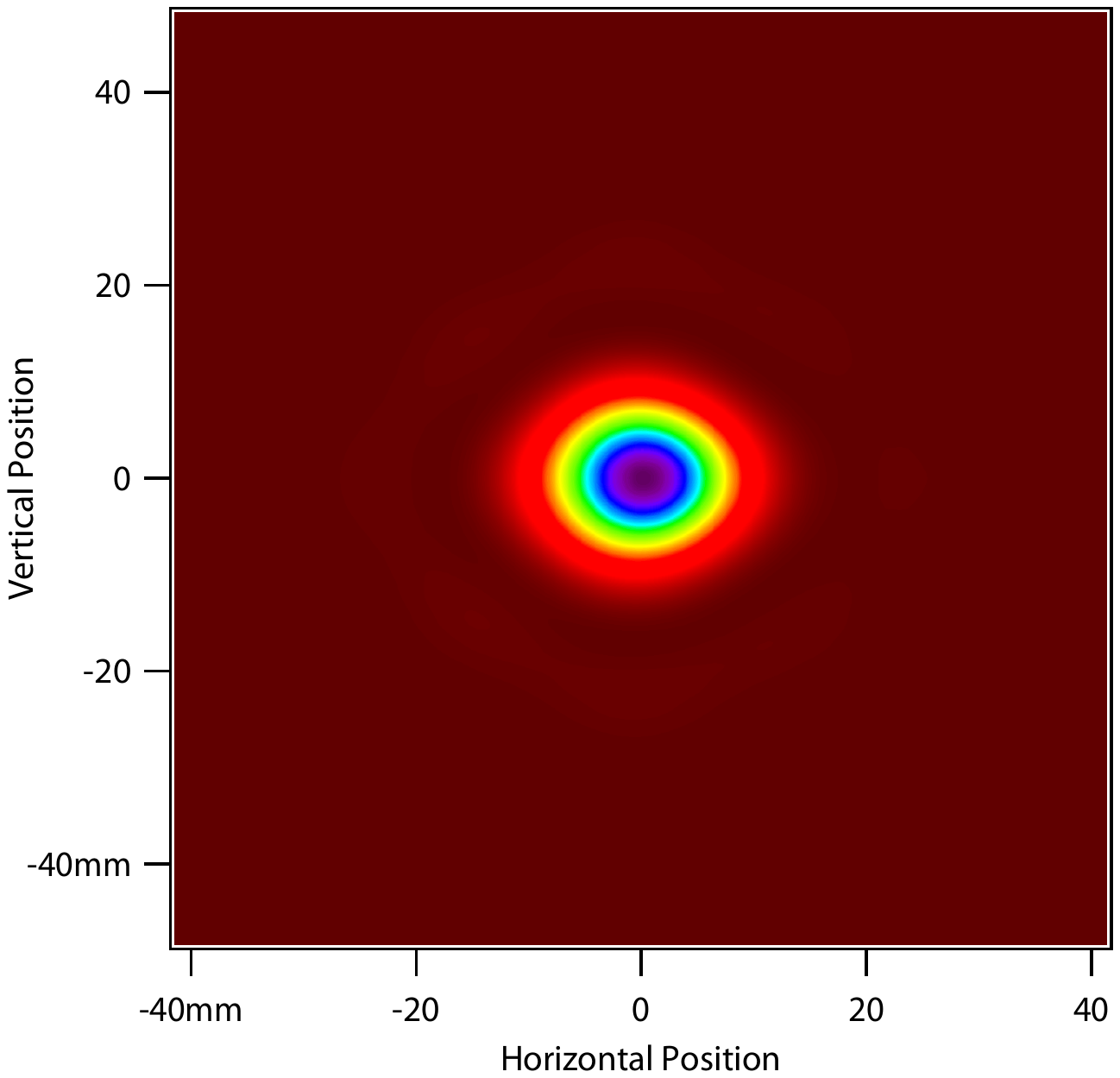}
\includegraphics[trim = 0 0 0 0, clip, width=0.4\textwidth]{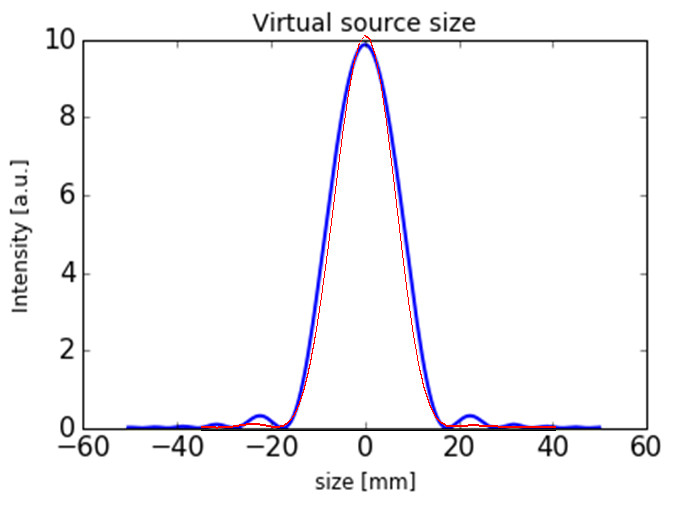}
\end{center}
\caption{SRW \cite{SRW} back propagation at the source for a fundamental of 3~THz with an electron beam energy of 17.5~GeV. The blue line in the lower plot is calculated taking the squared modulus of the field in Eq.~(\ref{undurad5}). The red line is the result of SRW calculations.} 
\label{fig:8}
\end{figure}

It is also possible to make numerical computations using wavefront propagation codes~\cite{CIT3}. We first calculated the radiation field generated by a filament beam at a distance of 25~m from the undulator center. Then, we backpropagated radiation in free-space up to the undulator center. The results of backpropagation at different distances and the calculation of the FWHM are shown in Fig.~\ref{fig:7}, where we also plot a comparison with the analytical expressions from Eq.~(\ref{undurad5}) and Eq.~(\ref{Esum}). Fig.~\ref{fig:8} shows the result of backpropagation at the virtual source. The blue line in the 1D cut is calculated taking the squared modulus of the field in Eq.~(\ref{undurad5}).

\section{\label{sect:THzSCU} Feasibility of the THz Superconducting Undulator}
The wavelength $\lambda$  of the radiation emitted at the first harmonic of a planar undulator with period length $\lambda_U$ and magnetic peak field $B$ is given in Eq.~(\ref{resonance}).

\begin{figure}
\begin{center}
\includegraphics[width=0.5\textwidth]{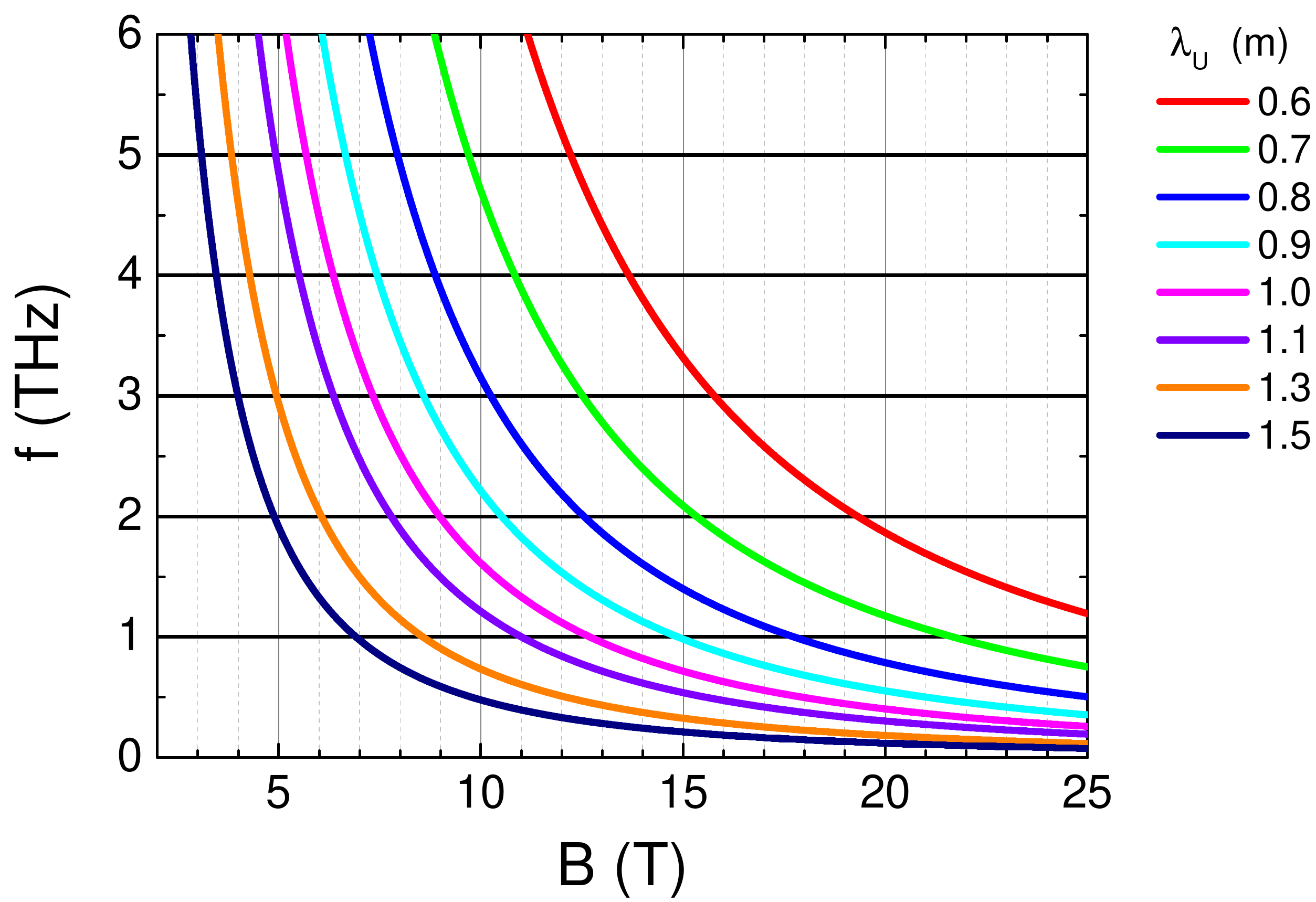}
\end{center}
\caption{Frequency of the first harmonic on axis ($\theta =0$) produced by an undulator with different period lengths, as a function of the peak field on axis B, for  17.5~GeV electron beam energy.}
\label{fversB}
\end{figure}

In Fig.~\ref{fversB} we report the frequency in THz of the first harmonic on axis ($\theta =0$) produced by an undulator with different period lengths, ranging from 0.6~m to 1.5~m, as a function of the peak field on axis B, for 17.5~GeV, the maximum nominal electron beam energy of the European XFEL. In order to reach 3~THz as the minimum value for the frequency of the first harmonic, for magnetic peak fields below 10~T the period length of the undulator should be longer than 0.9~m. To have some margin, and to keep the period length as short as possible we focus on a period length of 1~m.

Undulators are realized with different technologies~\cite{Clarke}. The most widely used is the permanent magnet technology. Alternatively, it is possible to use electromagnets wound with copper conductors or with superconductors. If the engineering current density needed to produce the desired magnetic peak field on axis is below 10~A/mm$^{2}$, electromagnets using copper as a conductor can be used. Above this limit of the engineering current density, the ohmic losses become untreatable. For the same period length and vacuum gap, especially for long period lengths, superconducting undulators can produce higher magnetic peak fields on axis than the permanent magnet ones~\cite{casalbuoni2014}. Superconducting undulators can also be used to obtain the same magnetic peak field using a larger vacuum gap, which allows to reduce the frequency cut off of the vacuum chamber, and to transmit most of the THz radiation of interest, starting from the first harmonic up to higher frequencies. Even using magnetic gaps down to 10~mm and 1.5~m period length, with permanent magnet technology it is not possible to reach peak magnetic fields on axis up to 4~T, as needed to obtain 3~THz (see Fig.~\ref{fversB}).

The superconducting technology to be chosen depends on the maximum field on the conductor necessary to reach the wished magnetic peak field on axis, on the operating temperature, and on the current density required.
The only superconductor used for accelerator magnets up to now is NbTi, whose radiation hardness has been widely proven in different accelerators over many years of operation, see for example the Tevatron~\cite{Tevatron}, HERA~\cite{HERA}, and the LHC~\cite{LHC}.
Superconducting wigglers based on NbTi technology with a closed loop liquid helium circuit and the coils in liquid helium bath, with period lengths from 30~mm to 148~mm, developed at the Budker Institute of Nuclear Physics (BINP) are used worldwide in many third-generation synchrotron light sources~\cite{mezwal2011}.
A closed thermosyphon loop, including a liquid helium tank, is used to cool down the two NbTi superconducting undulators with a period length of 18~mm (and 16~mm) and 1.1~m (and 0.3~m) magnetic length, constantly running in the Advanced Photon Source at the Argonne National Laboratory~\cite{YuriIPAC17}.
A cryogen-free full-scale superconducting undulator based on NbTi technology with 15~mm period length, using only cryocoolers with no need of liquid helium and nitrogen, has been successfully tested for one year at the ANKA synchrotron light source at the Karlsruhe Institute of Technology (KIT)~\cite{casalbuoni2016}. A similar full-scale superconducting undulator with 20~mm period length is now installed in the electron storage ring KARA of the KIT light source~\cite{IOP17}.
With all this in mind, we consider a NbTi based THz superconducting undulator working at liquid helium temperature (4~K), feasible for the European XFEL.
NbTi operates up to magnetic fields on the conductor of about 10~T (see for example Ref.~\cite{Lee}). The field on the conductor depends on the magnet design, for example on the geometry of the winding and on the presence and geometry of iron. It is however higher or of the same order of the peak field on axis.

\begin{figure}
\begin{center}
\includegraphics[width=0.4\textwidth]{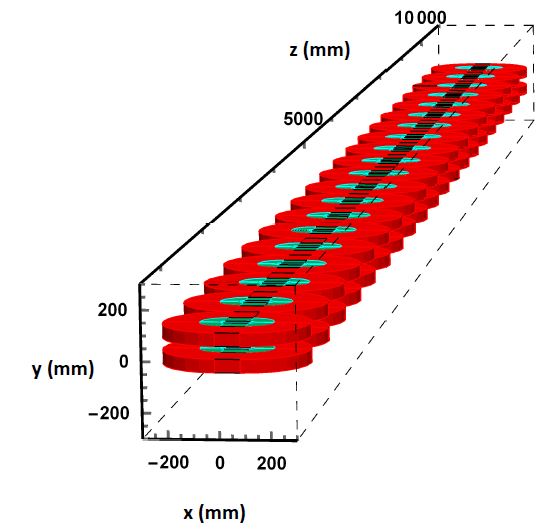}
\end{center}
\caption{Ten periods of the THz undulator model simulated with the code Radia. Green parts show the inner coils and red parts show the outer coils.}
\label{radiamodel}
\end{figure}

\begin{figure}
\begin{center}
\includegraphics[width=0.35\textwidth]{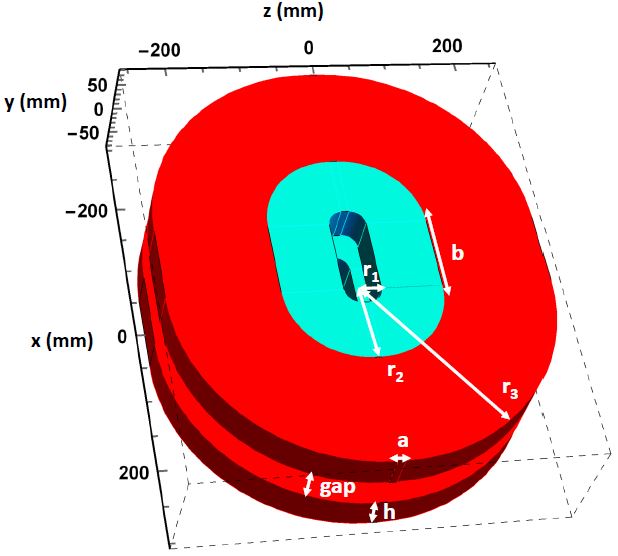}
\end{center}
\caption{Coil geometry of one pole. The straight lengths of the inner and outer coils are a = 10~mm and b = 100~mm, the inner coil bend radii are r$_1$ = 19.5~mm and r$_2$ = 100~mm, the outer coil bend radii are r$_2$ = 100~mm and r$_3$ = 245~mm. The coils height is h= 53.5~mm, while the magnetic gap is 50~mm.}
\label{geometrycoils}
\end{figure}

A possible solution, similar to what is used in the BINP superconducting wigglers~\cite{mezwal2011}, with the only difference that it does not foresee iron, is the Radia model~\cite{radia} shown in Fig.~\ref{radiamodel}. One pole consists of two sets of horizontally wound racetrack coils with the dimensions indicated in Fig.~\ref{geometrycoils}.
By choosing the cross section of the inner coil double as the one of the outer coil, it is possible to power the two sets of coils in series, making use of only one power supply and of one set of current leads, reducing the heat intake with respect to a configuration with two sets of current leads.

\begin{figure}
\begin{center}
\includegraphics[width=0.5\textwidth]{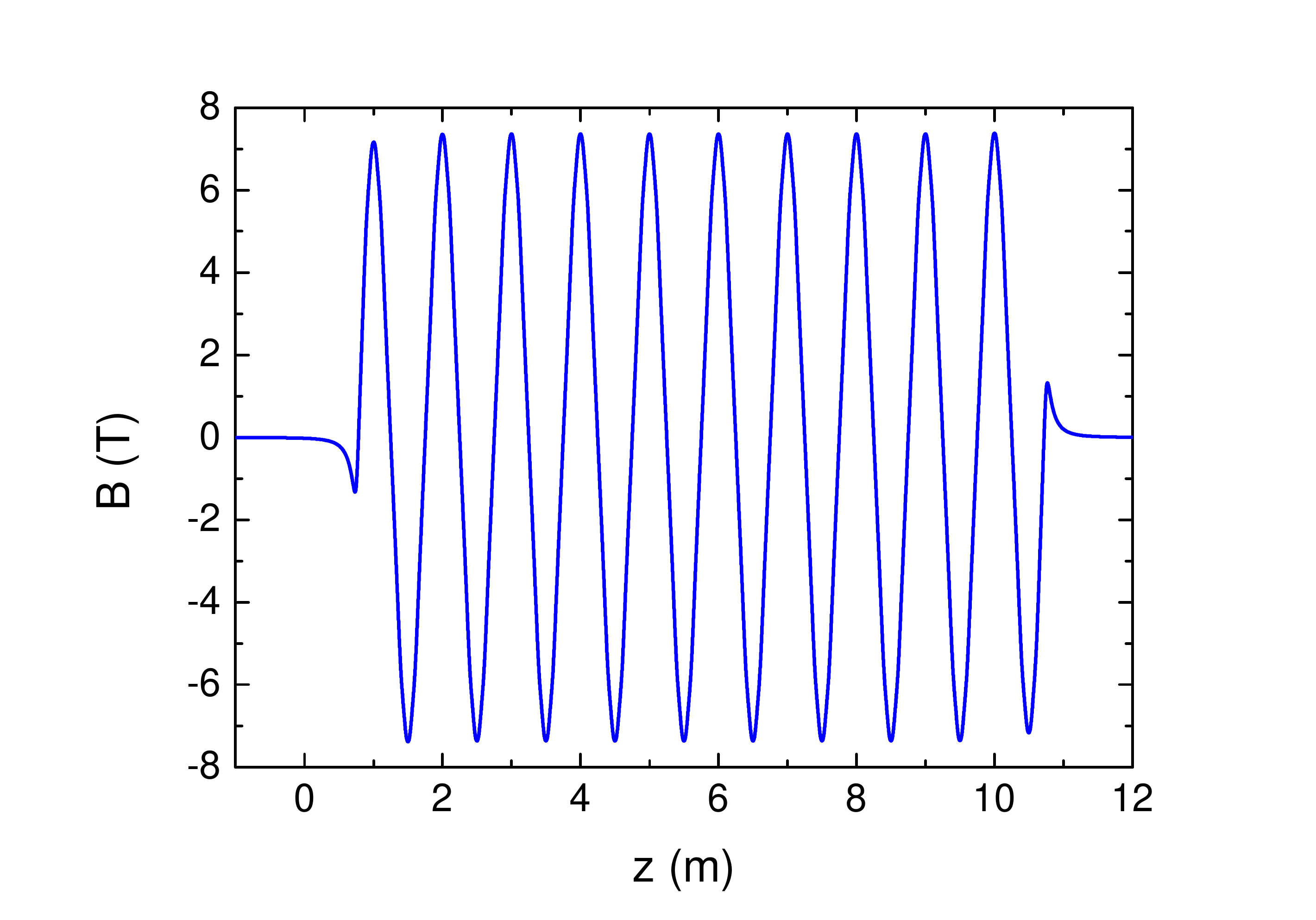}
\end{center}
\caption{Magnetic field profile along the magnetic axis.}
\label{field}
\end{figure}

The magnetic field profile of a 10 periods undulator model from Fig.~\ref{radiamodel}, simulated with the code Radia~\cite{radia}, is shown in Fig.~\ref{field}. The simulation has been performed using the engineering current densities of 110~A/mm$^{2}$ in the outer coils, and 55~A/mm$^{2}$ in the inner coils, with the dimensions as indicated in Fig.~\ref{geometrycoils}.
The on-axis magnetic peak field of 7.3~T allows to reach the minimum value of 3~THz for the frequency of the first harmonic.

\begin{figure}
\begin{center}
\includegraphics[width=0.4\textwidth]{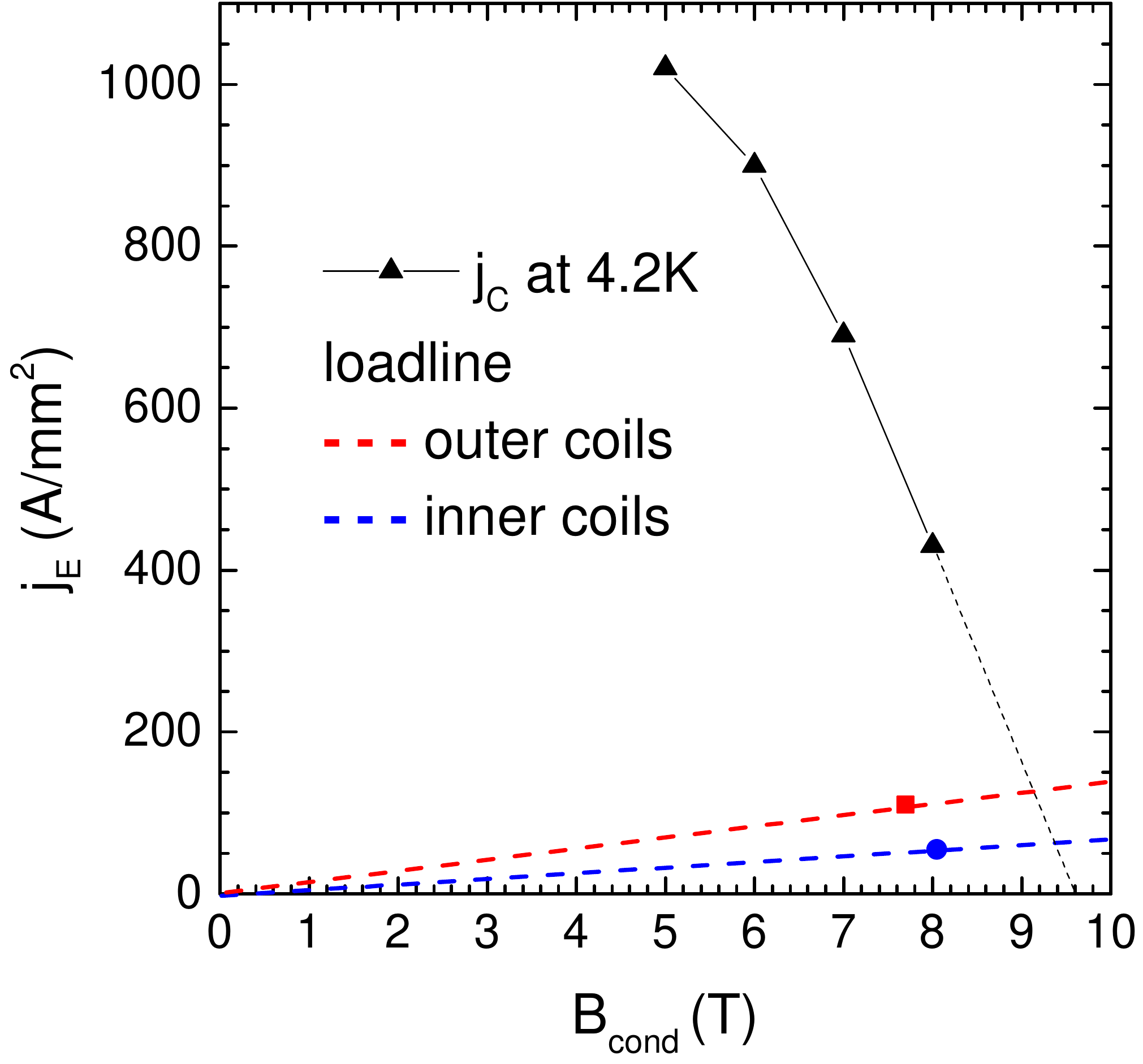}
\end{center}
\caption{Engineering critical current density of NbTi at 4.2~K~\cite{Lee} as a function of magnetic field (triangles), as well as the loadlines of the conductors in the inner and outer coils (dashed blue and red lines, respectively). The black dashed line indicates the linearly extrapolation to $j_E=0$~A/mm$^{2}$.}
\label{loadlines}
\end{figure}

The maximum magnetic peak field on the conductor is 8.0~T on the inner coils and 7.7~T on the outer coils. With these current densities and maximum magnetic fields on the conductors, NbTi conductor at about 4~K can be used with a large margin of operation both in current as well as in temperature.
This is demonstrated in Fig.~\ref{loadlines}, where we report the engineering critical current density of NbTi at 4.2~K~\cite{Lee} as a function of magnetic field (triangles), as well as the load lines of the conductors in the inner and outer coils (dashed blue and red line, respectively). The red square and the blue dot are the operation points to reach the maximum peak field on axis.

A detailed design of the undulator together with the other components, like for instance correction coils, is not in the scope of this report. Here we discussed a 10~m-long undulator, which we can installed at the European XFEL either in the XTD4 tunnel or in XTD5 tunnel see e.g.~\cite{XTDR}, just in front of the beam dump.

\section{\label{sec:tre} Feasibility of the transport line}

\begin{figure}
\begin{center}
\includegraphics[width=0.5\textwidth]{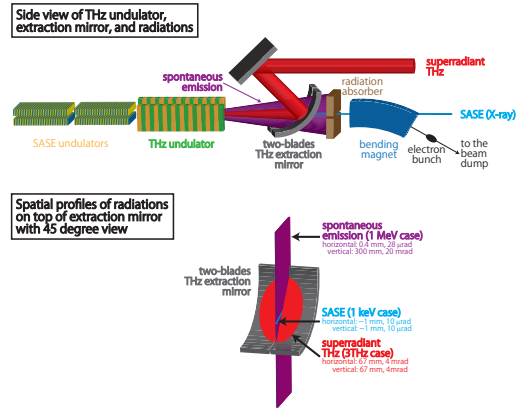}
\end{center}
\caption{The envisioned scenario of a superradiant THz undulator close to the electron beam dump with the undulator plan oriented vertically (top) and a schematic of the envisioned two-bladed outcoupling optics to circumvent damage due to the wide fan of parasitic spontaneous hard X-ray radiation.}
\label{THzpath}
\end{figure}

For the discussion of the feasibility of the transport of the generated THz pulses to the experimental endstations, we assume that the THz undulator would be positioned just in front of the electron beam dump as is the case at the prototype device at the XUV FEL FLASH~\cite{Gensch2008}. Then, the distances between the position of the THz undulator and the experimental end stations heavily depend on the facility. For example, for SACLA it is below 60~m~\cite{PRIV}, while at the European XFEL it is between 200 and 290~m. This poses the question of the feasibility of transport to the experiments. In the most cost-efficient configuration, as employed at the prototype facility at FLASH, the THz undulator is installed in the straight section following the SASE undulators. This has the advantage that the electron bunches do not need to be deflected and their pristine longitudinal form is fully preserved. THz and X-ray photon beams as well as the electron beam are in this case propagating collinearly after the THz undulator and the THz beam has to be appropriately separated. Outcoupling of the THz beam should occur as soon as possible after the THz undulator in order to avoid diffraction losses. The beam pipe diameter between the exit of the THz undulator and the outcoupling of the THz needs to be sufficiently large. One technologically feasible solution is to implement a metal optics with an appropriately large central aperture into the beamline that deflects the already wide THz beam out while letting the pencil-like X-ray and electron beam~\cite{NUMB} pass. One obvious obstacle is potential damage by the X-ray and/or electron beam on the metal optics. Experiences at the prototype facility FLASH, where an XUV beam with an average energy up to a few hundreds of $\mu$J traverses a central aperture of 10~mm diameter~\cite{Gensch2008} and at the second prototype facility TELBE~\cite{Green2016}, where a 30~MeV/10~$\mu$A electron beam is traversing a central aperture of 4~mm in diameter, show that this problem could be dealt with. In both facilities, no visible damage on the outcoupling optics has been observed despite years of user operation. At an XFEL, with its GeV class electron beam and keV X-ray photons,  this option would clearly require a fast interlock system to prevent damage in case of machine instabilities or failure. One additional aspect arises when implementing a superradiant THz undulator at a GeV class electron accelerator: the spontaneous hard X-ray radiation emitted into a relatively fan in the plane of the undulator. This shall be dealt with by utilizing a two blade design for the outcoupling optics as shown in the bottom part of Fig.~\ref{THzpath}.

In the standard case, the outcoupling optics would be an off-axis toroidal mirror that accepts a beam with the source point at the center of the THz undulator and thereby generates a collimated beam that is then deflected outside of the electron/X-ray beamline. The collimated THz beam would be refocused as soon as possible and optical transport could from thereon be performed via a series of appropriate Kepler telescopes next to the X-ray optical beamlines. This approach has been successfully employed at the FLASH and TELBE THz beamlines~\cite{Green2016, Gensch2008}. Optical transport should be performed in a vacuum of better than 0.1~mbar in order to avoid water vapor absorption. Experiences at FLASH and TELBE showed that keeping the accepted beam diameter by the optical elements to below 150~mm has some practical advantages. It ensures affordable metal optics and that a cost-efficient beamline diameter of 200~mm can be employed. Experience at FLASH furthermore shows that Kepler telescopes of a length of 12~m allow transport and refocusing without diffraction losses~\cite{Gensch2008} down to frequencies of 3~THz. A distance of 290~m, as needed in the case of the XTD4 tunnel at the European XFEL, would then require 24 Kepler-telescopes and a total of 48 mirrors (of which 32 would be focusing optics and 16 would be planar mirrors). The main loss will be due to the actual reflectivity. At FLASH and TELBE gold-coated aluminum or copper optics are employed, with the gold coating being a layer of about 200~nm, substantially thicker than the skin depth at these frequencies. Then the reflectivity of the optics can be as high as 99~$\%$~\cite{Naftaly2011} resulting in a reflectivity loss of 36~$\%$ for the mere transport to the experimental endstations. Additionally, there potentially will be losses in the window separating the accelerator vacuum from the experiment vacuum, which are in the few 10~$\%$ range, depending on the choice of material. Therefore roughly 50~$\%$ of the generated pulse energies are received at the experiments.

\section{\label{sec:quattro} Synchronization and jitter issues of pump and probe signals}
Matching the arrival time of X-ray and THz photon pulses, generated by the same accelerator, so that they arrive synchronously in the experimental end stations is not a trivial task. Due to the requirement of repeated focusing for the THz beam, the path length from the source to the experimental end stations for the X-ray pulses will be considerably shorter than that of the THz pulses. This path difference needs to be compensated in order to perform ultra-fast experiments based on the pump-probe technique. The exact path difference will depend on different specific boundary conditions at each XFEL facility. Assuming a standard design as employed at FLASH~\cite{Gensch2008}, where a series of Kepler telescope would be used for transport and subsequent refocusing of the THz beam, an additional 1~m of path length would arise per Kepler telescope. This would yield a mismatch between the X-ray and the THz optical path of the order of 24~m, making the THz pulse generated by the same electron bunch arrive roughly 80~ns too late at the European XFEL. Assuming a repetition rate of 4.5~MHz (micro pulse separation of 220~ns) the THz pulse from the previous electron bunch in the train would arrive 140~ns too early. In principle, there are now two all-optical approaches to circumvent this problem. Option one, chosen at the XUV FEL FLASH, is the compensation of the path difference by introducing an additional delay of the right length into the XUV path (the X-ray path in our case) so that XUV pulses and THz pulses from the same electron bunch can be used in the experiment~\cite{Gensch2008}. This has the advantage that these pulses are then intrinsically synchronized on the few femtosecond level~\cite{Fruehling2009} but has the drawback that a lot of flexibility is lost for the wavelength of the XUV pulses as specific multilayer optics have to be developed for each XUV wavelength of choice. In the harder X-ray regime, this approach would, if at all possible, require multiple grazing incidence optics. More feasible seems, therefore, to compensate for the path length difference by introducing an additional delay into the THz beam path and match the arrival time of a THz pulse arriving early to the X-ray pulse generated by the following electron bunch in the micro-pulse train. A cavity arrangement would be technologically feasible and allow generating an adjustable path length difference corresponding to a delay of about 140~ns. The overall length assuming the proposed folded symmetry would be below 12~m and hence acceptably sized. The additional reflection loss due to the 3 additional reflections would be negligible within a 3~$\%$ accuracy. The drawback of these arrangements that it is in this form only applicable for the highest repetition rate of the European XFEL of 4.5~MHz. Adjusting the proposed scheme to lower repetition rates (e.g. 2.25~MHz) would result in introducing multiple turns inside of the cavity which ultimately leads to higher reflectivity losses (additional 6 metal optics result in 6~$\%$ reflectivity loss).
One more complex alternative approach is a double pulse scheme by which an appropriately timed second electron bunch generates the THz pulse sufficiently in advance. This scheme would yield considerably higher flexibility and allow operation at any given repetition rate and has recently been discussed and successfully tested at the FLASH XUV FEL~\cite{Grimm2006, Stojanovic2017} aside for being used e.g. at the LCLS~\cite{Marinelli2015}. At a superconducting RF linac based on 1.3~GHz technology such a scheme allows to place electron bunches in multiples of 770~ps distance with respect to each other. A THz optical delay line of 1~ns would be hence sufficient to match the arrival time of THz and X-ray pulses in the experiment.
One remaining issue is the additional X-ray pulse that would be generated by this double electron bunch alongside the THz pulse. At the European XFEL, it would arrive at the experiment roughly 140~ns after the THz excitation and the X-ray probe pulse. Depending on the bandwidth of the detectors utilized, experiments may be able to gate this pulse out. Another approach would be to force the electron bunch producing the THz pulse along a non-ideal trajectory in the SASE undulators, enough to suppress the X-ray pulse emission in the undulator section.

The issue of the actual timing of the THz pulses to the X-rays has been technologically solved. Different approaches provide few femtosecond time-resolution (see e.g.~\cite{Tavella2011}) and these pulse-resolved detection schemes have recently been shown to be scalable from a few 100~kHz to 4.5~MHz repetition rates~\cite{Kovalev2017}.

\section{\label{sec:cinque} Conclusions}

\begin{table}
\caption{Maximum pulse energy (total, up to 100~THz and around the fundamental frequency with 10~$\%$ bandpass filter) at different values of the fundamental and corresponding electron charge (see Fig.~\ref{fig:6b}).}
\begin{tabular}{c c c c}
\hline
              Fund.      &           Tot. pulse   &    Fund. pulse      &           Electron              \\
              frequency [THz]              &           energy [$\mu$J]    &     energy [$\mu$J]   &           charge [pC]              \\
              \hline
  2.8       &           3450    &    279        &           500        \\
  4.8       &           2600    &    172        &           500        \\
  6.6       &           1540    &    115        &           250        \\
  8.5       &           1340    &    84.0       &           250        \\
  10.4     &           1180    &    63.8       &           250        \\
  12.3     &           1050    &    44.5       &           250        \\
  14.3     &           955    &    31.4       &           250        \\
  19        &           441    &    24.7       &           100        \\
  23.5     &           388    &    20.1       &           100        \\
  28.5     &           345    &    14.1       &           100        \\
  33        &           311    &    11.1       &           100        \\
  38        &           285    &    10.2       &           100        \\
  42.5     &           263    &    9.6         &           100        \\
  47.5     &           245    &    8            &           100        \\
\hline
\end{tabular}
\label{table:fig44}
\end{table}

In this paper, we explored the possibility of using the spent electron beam at the European XFEL to produce powerful pulses of superradiant THz radiation by means of a superconducting undulator. We show that current superconducting technology allows for the construction of such device with 10-periods with fundamental frequency starting from from 3~THz using a 17.5~GeV electron beam. Lower frequencies can be obtained by using lower nominal electron energy points. Owing to the quality of XFEL electron beams, one can generate narrow-band, frequency-tunable THz pulses with the frequency range of 30 down to 3~THz and corresponding pulse energies range of 10 to 280~$\mu$J (see Table~\ref{table:fig44}). Furthermore, mJ-level pulse energies are achievable when the full bandwidth between 1 and 100~THz is utilized. The repetition rate naturally follows that of the European XFEL. Even though the natural synchronization between the X-ray pulses and the THz pulses obtained from the same electron bunch cannot be exploited, we show that synchronization and jitter issues are actually not a problem in achieving pump-probe capabilities. Moreover, the THz transport line, albeit several hundred meters long, is feasible and can be hosted in the tunnels designed for the X-ray transport.

\section*{Acknowledgements}
The authors thank Serguei Molodtsov, Joachim Pflueger and Thomas Tschentscher (all from European XFEL) for their support and interest in this work, and Nikola Stojanovich (DESY) for many useful discussions. Vivek Asgekar acknowledges support from the Alexander von Humboldt Foundation, Germany. Sergey Kovalev and Michael Gensch acknowledge support from the European Cluster of Advanced Laser Light Sources (EUCALL) project, which has received funding from the European Union’s Horizon 2020 research and innovation programme under grant agreement No.~654220.


\begin{thebibliography}{99}
\bibitem{NotaL} While we refer to the LCLS-II to point out challenges related with high-repetition rate, we do not imply that our feasibility study should be adapted to this particular facility.
\bibitem{CIT1} In this paper we will refer to the optical frequency as $\nu = c/\lambda$, where $c$ is the speed of light and $\lambda$ the radiation wavelength.
\bibitem{CIT2b} It should be noted that at $8.5$~GeV one would be able to reach into lower photon energies with the same technology. In this case, about $3.6$~T, corresponding to $K=338$ are needed to reach 3~THz, and therefore the same technology can be used.
\bibitem{CGS} In this paper, unless otherwise stated, we will use the Gaussian unit system.
\bibitem{NotaT} Here we neglect any dependence on the transverse electron beam properties because we discuss about a class of electron beams suitable for hard X-ray FELs that are, therefore, diffraction limited up to much higher frequencies that terahertz.
\bibitem{noBW} In case of none and 200~mm apertures, the angular filtering is not sufficient to resolve the fundamental spectrum.
\bibitem{CIT3} In our case we used SRW~\cite{SRW}, which only relies on the use of the paraxial approximation.
\end{thebibliography}
\end{document}